\documentstyle[aps,prl,epsf]{revtex}
\begin{document}
\twocolumn[\hsize\textwidth\columnwidth
\hsize\csname@twocolumnfalse\endcsname
\draft

\title{Micro-SQUID technique for studying the temperature 
dependence of switching fields of single nanoparticles}

\author{\textnormal{C. Thirion$^{1}$, W. Wernsdorfer$^{1}$, M. Jamet$^{2}$, 
V. Dupuis$^{2}$, P. M\'elinon$^{2}$, A. P\'erez$^{2}$, D. Mailly$^{3}$}}

\address{
$^{1}$ Laboratoire Louis N\'eel-CNRS, BP 166,
        38042 Grenoble, France.\\
$^{2}$ DPM-Universit\'e Claude Bernard-Lyon 1 et CNRS,
    69622 Villeurbanne, France.\\ 
$^{3}$ L2M-CNRS,
        196 avenue H. Ravera, 92220 Bagneux, France.
}
\date{{\bf Paper number: F024.} Version: \today }
\maketitle

\begin{abstract}
An improved micro-SQUID technique is presented allowing us to
measure the temperature dependence of the magnetisation switching fields 
of single nanoparticles well above the critical superconducting 
temperature of the SQUID.
Our first measurements on 3 nm cobalt nanoparticle
embedded in a niobium matrix are compared to the N\'eel Brown
model describing the magnetisation reversal by thermal activation
over a single anisotropy barrier.
\end{abstract}

\bigskip
\pacs{\bf Keywords: Stoner-Wohlfarth model, Thermal activation, 
Nanoparticle, SQUID, Magnetometry}
\vskip1pc]
\narrowtext

In order to avoid the complications due to 
distributions of particles sizes, orientations, etc., 
which are always present in assemblies of particles~\cite{Dormann97}, 
single-particle measuring techniques have been developed 
such as magnetic force microscopy~\cite{Ledermann94} 
or magnetometry based on micro-Hall probes~\cite{Geim98}. 
Another powerful measurement technique is based on micro-SQUID devices.
In this paper we review the an improved micro-SQUID set-up
which allow us to detect the magnetisation reversal of 
single nanoparticles. In respect to previous studies~\cite{WWPRL97}, we
improved the sensitivity of the micro-SQUID technique by more
than two orders of magnitude. Furthermore, we developed a
technique to measure the switching fields for all 
direction of the applied field.
Finally, we extended the
temperature range to the interval from 30 mK to 30 K.
We present these improvements in the following.

\begin{figure}
\centerline{\epsfxsize=7 cm \epsfbox{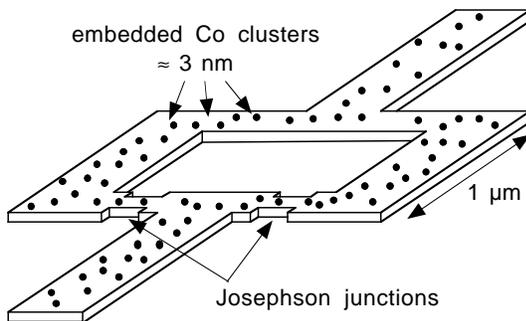}}
\caption{Schematic drawing of a micro-bridge-DC-SQUID which is patterned
out of a 20 nm-thick superconducting niobium film containing a low density
of 3 nm cobalt nanoparticles (black dots). The concentration is low enough ($<$
0.1 $\%$) in order to have no more than 5 particles located in a micro-bridge
(300 $\times$ 50 nm$^{2}$). The magnetic flux coupling of only the nanoparticles
in the micro-bridges was strong enough to give a measurable signal for each
individual nanoparticle. This new configuration detects the magnetisation
reversal of few hundred of spins.}
\label{SQUID}
\end{figure}

In order to measure the magnetisation reversal of a single 3 nm
Co nanoparticle (containing about 1000 atoms), we have to couple
the magnetic stray field of the nanoparticle with the SQUID loop.
In previous studies on larger particles (around 10 nm or larger), 
it was possible to place the nanoparticle close to 
the micro-bridge Josephson junctions of the SQUID~\cite{WWPRL97}.
This is not sufficient for 3 nm Co nanoparticles due to the small size
of the nanoparticle compared with the micro-bridge. Moreover, it
is very difficult to protect them efficiently against oxidation.
We solved these problems by directly embedding the Co nanoparticles
into the micro-bridges (Fig. \ref{SQUID}). This was achieved by using
a laser vaporisation and inert gas condensation source
producing an intense supersonic beam 
of nanosized Co nanoparticles under 
ultra-high-vacuum (UHV) conditions. 
A niobium matrix was simultaneously deposited
from a UHV electron gun evaporator
leading to continuous films (typical 20 nm thick) with a very low 
concentration of embedded Co nanoparticles~\cite{JametPRB00}.
These films are used to pattern planar microbridge-DC-SQUIDs
by electron beam lithography.
The desired sensitivity is only achieved for Co-nanoparticles embedded
into the micro-bridges where the magnetic flux coupling is high enough.
Due to the low concentration of embedded Co-nanoparticles, we have a maximum of
5 non-interacting particles in a micro-bridge 
(300 nm long and 50 nm wide). We can
separately detect the magnetic signal for each nanoparticle. Indeed they are
clearly different in intensity and orientation because of the random
distribution of the easy  magnetisation directions.

The switching field of the magnetisation of a single Co nanoparticle
can be measured by using the {\it cold mode} method described in 
detail in Ref.~\cite{WWJAP00}. However, this mode works only in
the low temperature regime and for magnetic fields applied
in the SQUID plane. We therefore developed the {\it blind mode}
method~\cite{Bonet99} allowing us to measure the switching fields for all 
directions of the applied field by separately 
driving three orthogonal coils.
In addition, the {\it blind mode} allowed us to extend the
temperature interval from 30 mK to 30 K~\cite{JametPRL01}, i.e.
to temperatures well above the critical superconducting 
temperature of the micro-SQUID.

\begin{figure}
\centerline{\epsfxsize=7 cm \epsfbox{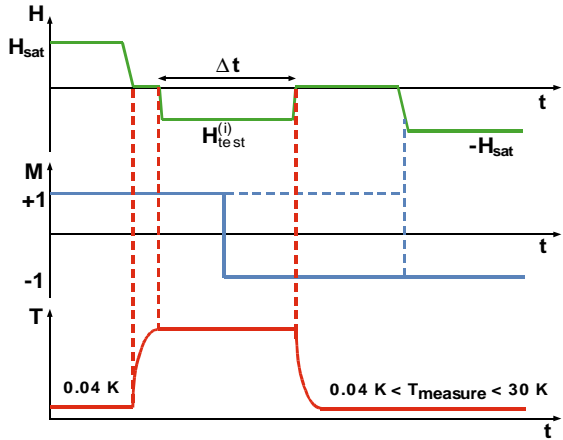}}
\caption{Chronogram of the {\it blind mode} method. The applied
magnetic field, the magnetisation state of the nanoparticle,
and its temperature are represented as a function of time.}
\label{chronogram}
\end{figure}

Let us consider Fig.~\ref{chronogram} showing schematically
the three steps of the {\it blind mode} method. The applied
magnetic field, the magnetisation state of the nanoparticle,
and its temperature are represented as a function of time.

\begin{enumerate}
\item
{\bf Saturation.} The magnetisation of the particle is saturated
in a given direction (at T = 35 mK).

\item
{\bf Testing.} A test field is applied at a temperature between 35 mK
and 30 K and during the time $\delta t$ 
which may or may not cause a magnetisation switching. 

\item
{\bf Probing.} After cooling to $T$ = 35 mK, the SQUID is switched on
and a field is swept in the plane of the SQUID to probe
the resulting magnetisation state using the cold mode.
\end{enumerate}

If the SQUID detects a magnetisation jump in step (3), 
this means that the previously applied test-field was  
weaker than the switching field for the direction being 
probed in step (2). 
The next iteration will then be done with a stronger test field. 
On the other hand, if the SQUID does not detect any 
magnetisation jump in step (3), this means that the 
reversal occurred during step (2). The next interation will 
then be done with a weaker field.
When choosing the new test field with the help of a 
bisection algorithm, we needed about eight repetitions of the 
three steps in order to get the switching field with good precision.
This method allows us to scan the entire field space.

\begin{figure}
\centerline{\epsfxsize=7 cm \epsfbox{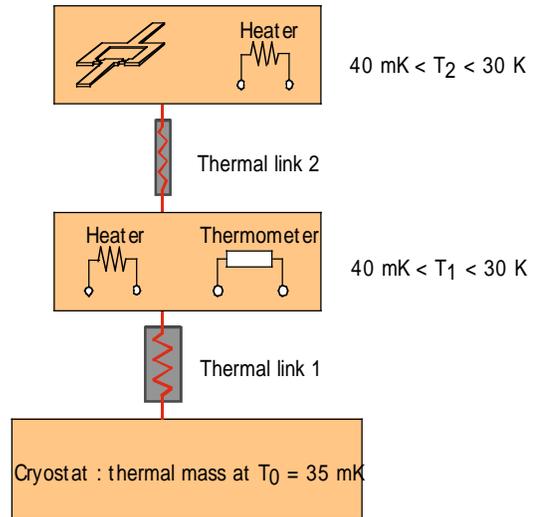}}
\caption{Two-stage heater which is used for the blind mode method.}
\label{heater}
\end{figure}

In order to probe the magnetisation reversal at high
temperatures, we have developed a two-stage heater 
shown in the Fig.~\ref{heater}.
The upper part of this device is a Si chip containing 
micro-SQUIDs and a resistor which is used as heating element.
The thermal response of this first stage is very fast because
it has a small thermal mass. This stage 
is coupled to a second stage with a much larger 
thermal mass. It contains another heating element and a thermometer.
Finally, the second stage is coupled to the cryostat
which has the largest thermal mass. The heat links 
between the two stages are metallic wires
which are adapted so that each stage is a thermal mass 
for its upper part. The cryostat is the 
thermal mass of the whole system and stays cold during
operation. The purpose of the second stage is to calibrate
efficiently the heater of the first stage.
With this set-up, our highest temperatures of 30 K 
was only limited by the cooling time.
Below 30 K, we achieved cooling rates of few
kelvins per second.

\begin{figure}
\centerline{\epsfxsize=7 cm \epsfbox{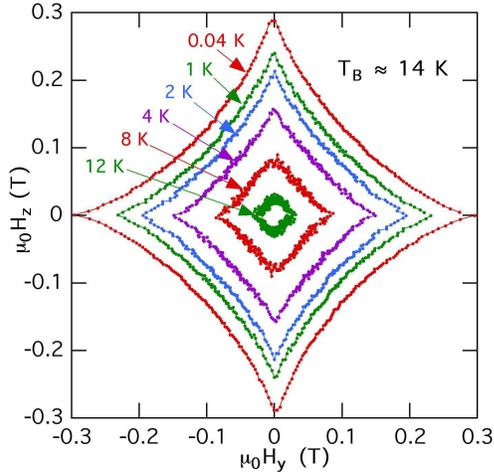}}
\caption{Temperature dependence of the switching field 
of a 3 nm Co nanoparticle, measured in the
plane defined by the easy and medium hard axes.
The data were recorded using the blind mode method
with a waiting time of the applied field of $\Delta t$ = 0.1 s.
The scattering of the data is due to stochastic.}
\label{ast_T}
\end{figure}

Fig.~\ref{ast_T} presents the angular dependence of the switching field of 
a 3 nm Co nanoparticle measured at different temperatures 
using the blind mode method. At 0.03 K, the 
measurement is very close to the standard Stoner--Wohlfarth astroid. 
For higher temperatures the switching field becomes 
smaller and smaller. It reaches the origin at about 14 K, yielding the 
blocking temperature $T_{\rm B} = 14$ K of the nanoparticle magnetisation. 
$T_{\rm B}$ is defined as the temperature
for which the waiting time $\Delta t$ becomes equal to
the relaxation time $\tau$ of the particle's magnetisation at
$\vec{H} = \vec{0}$. $T_{\rm B}$ can be used to estimate the
total number $N_{tot}$ of magnetic Co atoms in the nanoparticle. 
The N\'eel Brown model leads to
 $\Delta t = \tau = \tau_0 \exp(K_{at}N_{tot}/k_{\rm B}T_{\rm B})$,
where $\tau_0^{-1}$ is the attempt frequency typically
between $10^{10}$ to $10^{11}$ Hz, 
K$_{at}$ is an effective anisotropy energy per atom 
and $k_{B}$ is the Boltzmann constant. 
Using the expression of the
switching field at $T$ = 0 K and for $\theta = 0$:
$\mu_{0}H_{sw} = 2K_{at}/\mu_{at}= 0.3 T$ (Fig.~\ref{ast_T}),
the atomic moment $\mu_{at}=1.7\mu_{B}$,
$\Delta t$ = 0.01 s,
$\tau_0 = 10^{-10}$ s, and $T_{\rm B} = 14$ K,
we deduce $N_{tot} \approx 1500$, which
corresponds very well to a 3 nm Co nanoparticle.

In conclusion, we have shown that the micro-SQUID technique combined with
the Low Energy nanoparticle Beam Deposition is a powerful method to investigate
the magnetic properties of nanosized magnetic particles. In particular, it
allows to measure in three dimensions the switching field of individual
grains giving access to its magnetic anisotropy. Furthermore, the
temperature dependence of the switching field is measurable and allows to
probe the magnetisation dynamics.

We thank B. Barbara for his constant support of this research.
Part of this work has been
supported by DRET, Rh\^one-Alpes, and the MASSDOTS ESPRIT
LTR-Project $\#$32464.

\end{document}